# Spin-polarized chiral $ZnIn_2S_4$ for targeted solar-driven $CO_2$ reduction to acetic acid

Yongping Cui[1†], Yuanbo Li[2†], Zhi-qiang Wang[3†], Xueliang Zhang[2], Lu Han[2*], Xueli Wang[4], Jinquan Chen[4*], Aokun Liu[5,6], Lu Yu[5], Changlin Tian[5,6,7*], Xue-qing Gong[1*], Wanning Zhang[1] & Yuxi Fang[1*]

[1]State Key Laboratory of Synergistic Chem-Bio Synthesis, School of Chemistry and Chemical Engineering, Frontiers Science Center for Transformative Molecules, Shanghai Key Laboratory for Molecular Engineering of Chiral Drugs, Shanghai Jiao Tong University, 800 Dongchuan Road, Shanghai, 200240, China. [2] School of Chemical Science and Engineering, Tongji University, 1239 Siping Road, Shanghai, 200092, China. [3] State Key Laboratory of Green Chemical Engineering and Industrial Catalysis, Center for Computational Chemistry and Research Institute of Industrial Catalysis, School of Chemistry and Molecular Engineering, East China University of Science and Technology, 130 Meilong Road, Shanghai, 200237, China. [4]State Key Laboratory of Precision Spectroscopy, East China Normal University, Shanghai, 200241, China. [5]High Magnetic Field Laboratory, Hefei Institutes of Physical Science, Chinese Academy of Sciences, 350 Shushanhu Road, Hefei, 230031, China. [6]Division of Life Sciences and Medicine, University of Science and Technology of China, 96 Jinzhai Road, Hefei, 230026, China. [7]School of Chemistry and Chemical Engineering, Zhangjiang Institute for Advanced Sciences, Shanghai Jiao Tong University, 1308 KeYuan Road, Shanghai, 201210, China.

[†]These authors contributed equally to this work. *e-mail: sjtu15901600323@sjtu.edu.cn; luhan@tongji.edu.cn; jqchen@lps.ecnu.edu.cn; cltian@sjtu.edu.cn, xqgong@sjtu.edu.cn

**Acetic acid, an important industrial chemical, is a key target product for $CO_2$ reduction due to its dual role in carbon utilization and chemical feedstock supply[1-6]. Although photocatalytic $CO_2$ reduction (PCCR) can generate acetic acid alongside other multicarbon products, its yield is typically low[7-9], limited by competing reactions and inefficient C-C coupling. Herein, we report a chiral mesostructured $ZnIn_2S_4$ (CMZI) photocatalyst that achieves a remarkable acetic acid yield of 962 μmol $g^{-1}$ $h^{-1}$ with a high selectivity of 97.3 %. This yield is ten times higher than the current highest reported value, while attaining state-of-the-art selectivity[10]. The remarkable productivity arises from synergistic effect between chiral structure and sulfur (S) sites of CMZI. Chirality-induced spin polarization in CMZI stabilizes the key triplet OCCO intermediate, significantly promoting C-C coupling efficiency. Theoretical calculations reveal that the S sites on {102} crystal facets of $ZnIn_2S_4$ exhibit thermodynamic and kinetic preferences for acetic acid formation. This work offers critical insights into catalytic strategies for $CO_2$ reduction toward the efficient and scalable synthesis of various multicarbon products.**

Chiral materials exhibit unique electronic properties where electron motion generates chirality-induced spin selectivity (CISS)[11] due to asymmetric spin-orbit coupling (SOC)[12,13]. Our previous work demonstrated that chiral mesostructured inorganic materials exhibit spin polarization through CISS, mediated by electron transport or electronic transitions[14,15]. Extending this foundation, we have developed chiral mesostructured catalysts for $CO_2$ reduction[16-18]. The chirality of these catalysts simultaneously promotes the formation of the key C-C coupling intermediate triplet OCCO ($^3$OCCO)[16-18] and stabilizes this intermediate throughout the reaction pathway[19] to efficiently produce multicarbon products.

Herein, our approach is to achieve efficient production of acetic acid from photocatalytic $CO_2$ reduction (PCCR) by synergistically promoting the formation of $^3$OCCO and chemisorbed $C_2$ intermediates through two key features of chiral mesostructured $ZnIn_2S_4$ (CMZI): chirality-induced spin polarization effects and sulfur (S) sites.

$ZnIn_2S_4$ was selected as an optimal catalyst due to its suitable bandgap (2.2-2.6 eV) for facilitating multicarbon product formation[20,21]. The valence band (VB), composed 3*p* orbitals of sulfur atoms, provides strong hole localization, while the conduction band (CB) exhibits superior reducibility compared to oxide analogues, promoting efficient $CO_2$ activation[22-24]. Currently, $ZnIn_2S_4$ is known to exclusively produces CO rather than multicarbon products in PCCR, due to the inherent thermodynamic and kinetic constraints that impede efficient C-C coupling on its surface.

In the pathway leading to multicarbon products during PCCR, the critical C-C coupling intermediate OCCO is known to exist in both singlet OCCO ($^1$OCCO) and $^3$OCCO states[25] (Supplementary Fig. 1)[15]. $^3$OCCO possesses unpaired electrons that occupy two degenerate Π orbitals in a neutral state. The $^3$OCCO is speculated to be generated and more stable in parallel aligned electron spin than in antiparallel aligned electron spin.

In CMZI, the CISS effect generates spin polarization *via* photoinduced electronic transitions, leading to the preferential accumulation of electrons with stable spin orientations at the surface (Supplementary Fig. 2). This would promote metastable $^3$OCCO formation by aligning parallel electrons opposite to surface spins *via* the *Pauli* exclusion principle[19,26]. Concurrently, the S site in $ZnIn_2S_4$ plays a critical role in selectively promoting the reaction pathway toward acetic acid formation by stabilizing the product through coordination with the unsaturated carbonyl oxygen in synergy with Zn sites. The strong affinity for key reaction intermediates facilitates multi-electron transfer processes, thereby enabling stepwise hydrogenation and C-C coupling for targeted acetic acid synthesis. Thus, the stabilization of the key $^3$OCCO and chemisorbed $C_2$ intermediates enhances acetic acid yield while suppressing $C_1$ byproduct formation. In contrast, achiral catalysts lacking spin polarization form unstable $^1$OCCO (ref. [27]), which readily dissociates into two CO molecules, hindering C-C coupling and favoring $C_1$ byproduct formation.

**Synthesis of CMZI**

CMZIs were synthesized *via* a hydrothermal reaction in the solution containing indium and zinc source, using L/D-cysteine (Cys) as both the sulfur source and a symmetry-breaking agent (Methods and Supplementary Fig. 3). Electrostatic interactions between Cys and $In^{3+}$/$Zn^{2+}$ ions drove asymmetric co-self-assembly, leading to chiral mesostructures within CMZI crystals. During the crystallization, $Zn^{2+}$ and $In^{3+}$ were simultaneously sulfided to form an ordered lattice. Organic components were removed by washing of the precipitates (Supplementary Fig. 4). Samples synthesized with L-, D-Cys and racemic-Cys (rac-Cys) were labeled as L-, D-CMZI and RMZI, respectively, while achiral mesostructured $ZnIn_2S_4$ (AMZI) was synthesized using thioacetamide (TAA) as the sulfur source.

**Structural characterization of CMZI**

Powder X-ray diffraction (XRD) patterns of antipodal CMZIs, RMZI, and AMZI (Fig. 1a) demonstrate that all four samples exhibit a trigonal phase (hexagonal setting) of $ZnIn_2S_4$ (JCPDS Card No. 65-2023) with space group $P\bar{3}m1$ (ref. [28]). The observed peak broadening is attributed to deformation and structural defects within the material.

Scanning electron microscopy (SEM) images show that L/D-CMZI and RMZI are composed of flower-like particles with average diameter of ~0.8 μm, which are hierarchically constructed from nanoplates with a thickness of ~13 nm (Fig. 2b and Supplementary Figs. 5 and 6). The diameter of AMZI flower-like particles is ~1.5 μm (Supplementary Fig. 7).

Electron tomography (ET) was employed to reconstruct the three-dimensional (3D) structure of CMZI using REal Space Iterative REconstruction (RESIRE) algorithm[29] from thirty-eight images obtained from -44.24° to +59.24° with 3° increments (Supplementary Fig. 8). The tomograms of L-CMZI (Figs. 1c-f) reveal that the sample consists of stacked nanoplates consistent with SEM observations. Side-view cross-sections (Figs. 1c and d) show intrinsically curvature in each nanoplate, with clockwise rotation as the cutting plane moves from front to back. Correspondingly, top-down cross-sections (Figs. 1e and f) display progressive clockwise rotation when sliced from bottom to top, indicating the nanoplates possess a left-handed helical twist (Fig. 1g). Selected-area electron diffraction (SAED) patterns along the [001] and [0$\bar{1}$0] axis of individual nanoplate show that each nanoplate is a single crystal, growing predominantly within the *ab*-plane and the [001] axis is perpendicular to the surface (Supplementary Fig. 9). However, the diffused diffraction intensities indicate the structural deformation. This structural feature can be also revealed in the high-resolution TEM (HRTEM) image and the corresponding Fourier diffractogram (FD) (Fig. 1h), showing a twisted morphology. The enlarged image (inset of Fig. 1h) shows that the exposed surface is (001) and the (102) plane corresponds to the edge of the nanoplate. Several defects and lattice mismatches were also observed, as indicated by white arrows. The structure of D-CMZI is similar to that of L-CMZI; lattice distortion was observed in RMZI and was absent in AMZI (Supplementary Figs. 5-7).

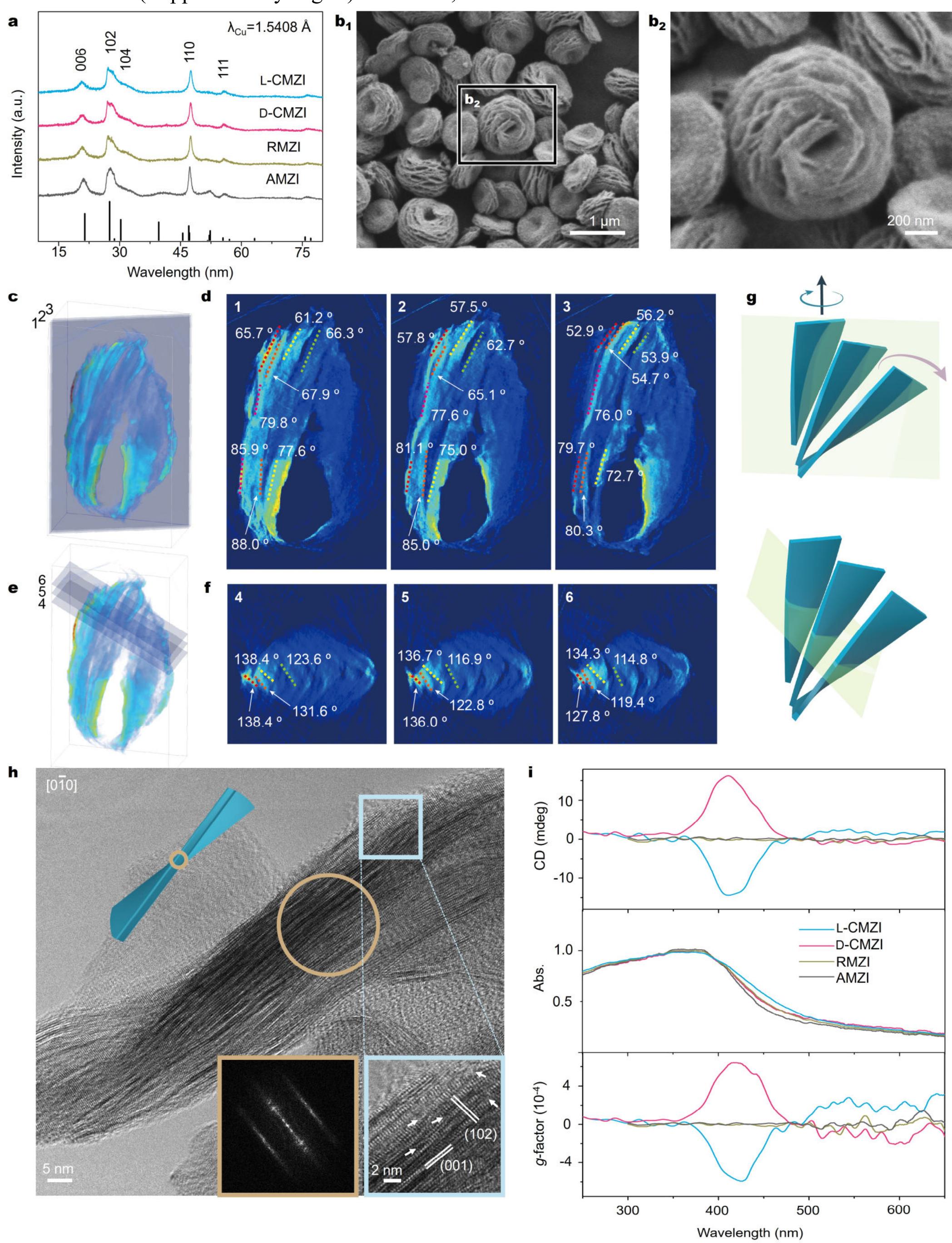

**Fig. 1 | Structure and morphology of CMZI. a,** Powder XRD patterns of antipodal CMZIs, RMZI and AMZI. Characteristic intensive diffraction peaks are observed at 2$\theta$ values of 21.6°, 27.7°, and 47.2°, corresponding to the 006, 102, and 110 crystallographic reflections of $ZnIn_2S_4$, respectively. **b$_{1,2}$**, SEM images of L-CMZI. **c**, Reconstructed 3D tomogram of L-CMZI with indication of position of slices 1-3. **d**, Cross sections of the tomogram viewing from side of the nanoplates. Angles of nanoplates identified were annotated. **e**, 3D tomogram of L-CMZI sliced from top direction (position 4-6). **f**, Cross sections of the tomogram viewing from bottom to top of the nanoplates. **g**, Schematic drawing of the chiral mesostructure. **h,** HRTEM image and corresponding FDs of one nanoplate in L-CMZI, taken along [0$\bar{1}$0] direction. **i,** DRUV-vis and DRCD spectra of antipodal CMZIs, RMZI and AMZI. The synthetic molar composition was: $Zn^{2+}$: $In^{3+}$: S: ethanolamine: $H_2O$ = 1: 2: 8: 48: 6,400.

The diffuse reflection circular dichroism (DRCD) spectra (Fig. 1i and Supplementary Fig. 10) of the antipodal CMZIs, exhibit mirror-imaged CD signals centered at ~410 nm, corresponding to absorbance of the electron transition from VB to CB, which validate their opposite handedness. The negative CD signal observed for L-CMZI indicates left-handed structure, and *vice versa* in D-CMZI, as the left-handed helical medium preferentially absorbs right-handed circularly polarized light (R-CP) and exhibits negative CD signals. The dissymmetry factor (*g*-factor) of antipodal CMZIs reaches a maximum value of $|g| \approx 5.3\times10^{-4}$ at a wavelength of 410 nm, providing a quantitative degree of chirality of the materials. As expected, RMZI and AMZI shows silent spectra due to positive and negative cancellations and the absence of chiral structure.

The high-resolution S 2*p* X-ray photoelectron spectroscopy (XPS) spectrum and electron paramagnetic resonance (EPR) confirmed the presence of similar sulfur vacancies ($V_S$) concentration in the antipodal CMZIs, RMZI and AMZI (Supplementary Figs. 11 and 12). The Brunauer-Emmett-Teller (BET) specific surface areas of L-CMZI, D-CMZI and RMZI (12.6, 14.7 and 13.5 $m^2\ g^{-1}$) are lower than that of AMZI (16.0 $m^2\ g^{-1}$) (Supplementary Fig. 13 and Table 1).

## Photocarrier behaviors of CMZI

UV-vis diffuse reflectance spectra (UV-vis DRS) of antipodal CMZIs, RMZI and AMZI (Fig. 2a) exhibit similar absorption edge at ~520 nm with comparable band gaps (2.37-2.40 eV), indicating nearly identical light absorption capabilities (Supplementary Fig. 14). The photoluminescence (PL) and femtosecond time-resolved transient absorption spectra (TAS) reveal enhanced charge separation in the antipodal CMZIs compared to RMZI and AMZI. Despite similar light absorption, CMZIs show weaker PL intensity at ~560 nm (Fig. 2b). Longer carrier lifetimes of CMZI could be observed from TAS (L-CMZI: 1,030 ps, D-CMZI: 996 ps vs RMZI: 575 ps, AMZI: 469 ps) (Supplementary Fig. 15 and Table 2), demonstrating superior charge separation efficiency.

Antipodal CMZIs exhibited 4-fold higher photocurrent (~0.85 μA $cm^{-2}$) than RMZI (~0.21 μA $cm^{-2}$) and AMZI (~0.20 μA $cm^{-2}$) (Fig. 2c), demonstrating significantly improved charge separation. The electrochemical impedance spectra (EIS) reveal lower charge-transfer resistance in the antipodal CMZIs compared to RMZI and AMZI (Supplementary Fig. 16), consistent with enhanced charge transport properties.

## Spin polarized electron transport of CMZI

Magnetic tip conducting atomic force microscopy (mc-AFM)[30] (Supplementary Figs. 17 and 18) reveals distinct spin-dependent conduction in L-CMZI: current exhibits a pronounced response to bias voltage under spin-UP magnetization but weakly under spin-DOWN (Fig. 2d), confirming preferential spin-UP electron transport, and *vice versa* in D-CMZI. The degrees of spin polarization (SP%)[31] of L- and D-CMZI are 79.3 and -75.2%, respectively. This CISS effect arises from chiral-structure-induced splitting of VB spin states, directing spin-UP (L-CMZI) or spin-DOWN (D-CMZI) electrons into the CB. In contrast, the control samples RMZI and AMZI exhibit negligible polarization (SP ≈ 0%), verifying that the chirality is essential for spin selectivity.

The dynamics of photoexcited charge carriers and spin states in CB of CMZIs were investigated using circularly polarized transient-absorption spectroscopy (CPTAS, Supplementary Fig. 19). Fig. 2e shows that the carrier lifetimes ($\tau_2$) probed with left-handed circularly polarized light (L-CP) is longer than that with R-CP probe, suggesting that spin-DOWN electrons are dominant in the CB of L-CMZI, and *vice versa* in D-CMZI (Fig. 2f and Supplementary Table 3). This finding implies that the predominantly excited spin-UP electrons observed in mc-AFM are inverted to the spin-DOWN state in L-CMZI, and *vice versa* in D-CMZI. As expected, $\tau_2$ of spin-DOWN and spin-UP electrons in the CB of RMZI and AMZI are almost the same (Supplementary Fig. 20).

The spin-polarization dynamics observed in the CMZI are driven by electron spin flips during charge transfer, a process that conserves spin angular momentum[11-13,32] (Supplementary Fig. 21). In L-CMZI, initial spin-UP electrons invert to spin-DOWN, while D-CMZI shows the opposite behavior. This spin filtering traps electrons in the CB, prolonging carrier lifetime (enhancing photocurrent) and enabling their participation in photocatalytic reactions.

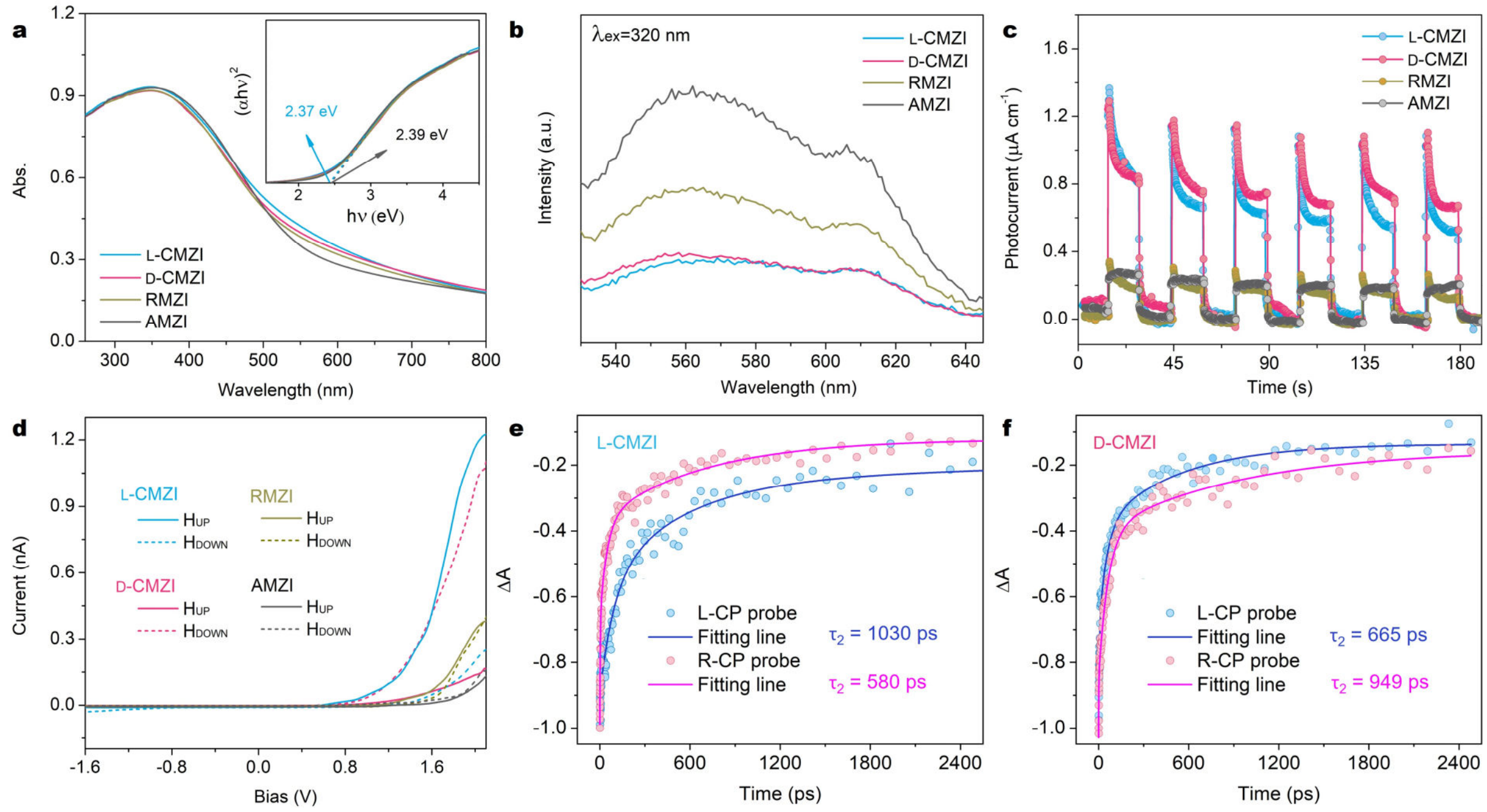

**Fig. 2 | Photoexcited charge separation properties and spin polarization of CMZI. a,** UV-vis DRS. Inset shows the Tauc plots obtained from DRS by plotting $[\alpha h\nu]^2$ versus h*v*, in which *α* and *v* are the absorbance coefficient and light frequency, respectively. **b,** Steady state PL spectra at an excitation wavelength of 320 nm. Small peak at ~610 nm could be ascribed to the scatter of exciting light. **c,** Time dependence of the photocurrent response under Xe lamp irradiation with an interval of 20 s. **d,** I-V curves obtained from mc-AFM measurements. **e, f,** CPTAS of antipodal CMZIs, measured with linear polarized pump and different circularly polarized probe beams at a probe wavelength of 360 nm.

## PCCR on CMZI

Catalytic performance was evaluated in a sealed quartz reactor in irradiation of 300 W Xe lamp equipped with an AM 1.5G filter (Methods). To exclude the effect of chiral ligand on the catalytic activity, L-Cys-AMZI (4.02 mol% L-Cys loading on AMZI) served as a control (Supplementary Fig. 22). Proton nuclear magnetic resonance ($^1$H NMR) and offline gas chromatography (GC) analyses identified $CH_3COOH$ as the primary liquid product (Supplementary Fig. 23a), $CH_4$ and CO as gaseous products (Supplementary Fig. 23b), respectively.

As shown in Figs. 3a and 3b (Supplementary Table 4), the L-CMZIs exhibit superior performance with a total product yield of 989 μmol $g^{-1}$ $h^{-1}$ and 97.3 % acetic acid selectivity. The yield of acetic acid reached up to 962 μmol $g^{-1}$ $h^{-1}$. In comparison, RMZI, AMZI and L-Cys-AMZI show markedly lower total yields (213, 162, 166 μmol $g^{-1}$ $h^{-1}$) and acetic acid selectivity (64.9, 30.3 and 33.4%, respectively). Notably, L-CMZI generate only 2.6% $C_1$ products (CO and $CH_4$), versus 35.1, 69.7 and 66.6% for RMZI, AMZI and L-Cys-AMZI, respectively. The comparable low performance of RMZI and AMZI demonstrates that catalytic activity is driven by chirality of CMZI. Control experiments confirm that neither chiral ligands (L-Cys-AMZI yields matching AMZI) nor non-catalytic conditions ($N_2$ atmosphere, catalyst-free, or dark conditions) produce significant products (Supplementary Fig. 24), verifying the photocatalytic origin of the observed activity in CMZI. The durability measurement proved the stability of the catalyst (Supplementary Fig. 25).

The apparent quantum yield (AQY) for acetic acid production reaches 4.36 % at 350 nm (Fig. 3c), consistent with L-CMZI's absorption spectrum, confirming the photocatalytic origin of $CO_2$ reduction (Supplementary Fig. 26). Isotopic labeling with $^{13}CO_2$, analyzed by gas chromatography-mass spectrometry (GC-MS), confirmed $CO_2$ as the carbon source, which was evidenced by $^{13}CH_3{}^{13}COOH$ ($m/z$=62), $^{13}CO$ ($m/z$=29) and $^{13}CH_4$ ($m/z$=17) products (Fig. 3d and Supplementary Fig. 27).

Fig. 3e presents a summary of the yield and selectivity of PCCR for acetic acid production[7-9, 33-38], as reported in the existing literatures (Supplementary Table 5). PCCR on CMZI achieved the yield ten times higher than the current highest reported value, while attaining state-of-the-art selectivity.

## Characterization of intermediates and products during PCCR

In situ FTIR spectra on L-CMZI (Fig. 4a), demonstrate $CO_2$ activation through characteristic adsorbate vibrations at 1,507 (m-$CO_3^{2-}$), and 1,274 $cm^{-1}$ (b-$CO_3^{2-}$). Key reaction intermediates include *CO (1,920 $cm^{-1}$), *HOCCO (1,572 $cm^{-1}$), and notably *OCCO (1,360 $cm^{-1}$), whose intensity growth under irradiation confirms its role as the dominant C-C coupling intermediate[16,39]. The formation of acetic acid is evidenced by *CHO vibrations (1,109 $cm^{-1}$) and *$CH_3COO^-$ stretch at 1,431 $cm^{-1}$ (ref.[36-38,40]).

The presence of $^3$OCCO was characterized by low-temperature EPR measurements at 2 K under a CO/$CO_2$ atmosphere under Xe lamp irradiation. L-CMZI revealed distinct paramagnetic signatures with g values at 1.996, 2.005, 2.006 and 2.029 (Supplementary Fig. 28). The $g$=2.006 signal corresponds to $CO^-$ (refs.[16-18,41]), while the additional peaks ($g$=1.996 and 2.029), which are absent in both RMZI and AMZI, originate from spin-polarization-induced species in L-CMZI. These signals are most plausibly assigned to the $^3$OCCO intermediate, suggesting preferential formation of $^3$OCCO on L-CMZI *versus* RMZI and AMZI.

## Spin polarized electron for the efficient C-C coupling

We propose a spin polarization-mediated PCCR pathway to acetic acid in CMZI (Fig. 4b). The mechanism involves three key steps: (i) *$CO_2$ → *CO conversion, (ii) *CO dimerization to form spin-polarized $^3$OCCO, stabilized by parallel spin alignment on CMZI 's surface[19,26,42], and (iii) sequential reduction of $^3$OCCO to *$OCCO^-$ (refs.[17,43,44]), *HOCCO, *$CH_2CO$, and finally *$CH_3COO$.

The observed ground-state electron spin polarization, excited-state population inversion, prolonged carrier lifetime, and the presence of the $^3$OCCO intermediate in CMZI collectively validate the mechanistic strategy proposed in the introduction. Under photoexcitation, the chiral structure of CMZI generates spin polarization *via* asymmetric SOC, which facilitates the formation of metastable $^3$OCCO through spin-aligned electron pairing, as dictated by the *Pauli* exclusion principle. The critical role of spin polarization is also evidenced by the inability of non-polarized catalysts (RMZI and AMZI) to stabilize $^3$OCCO or produce multicarbon products efficiently due to the absence of the parallel spin alignment required for intermediate stabilization.

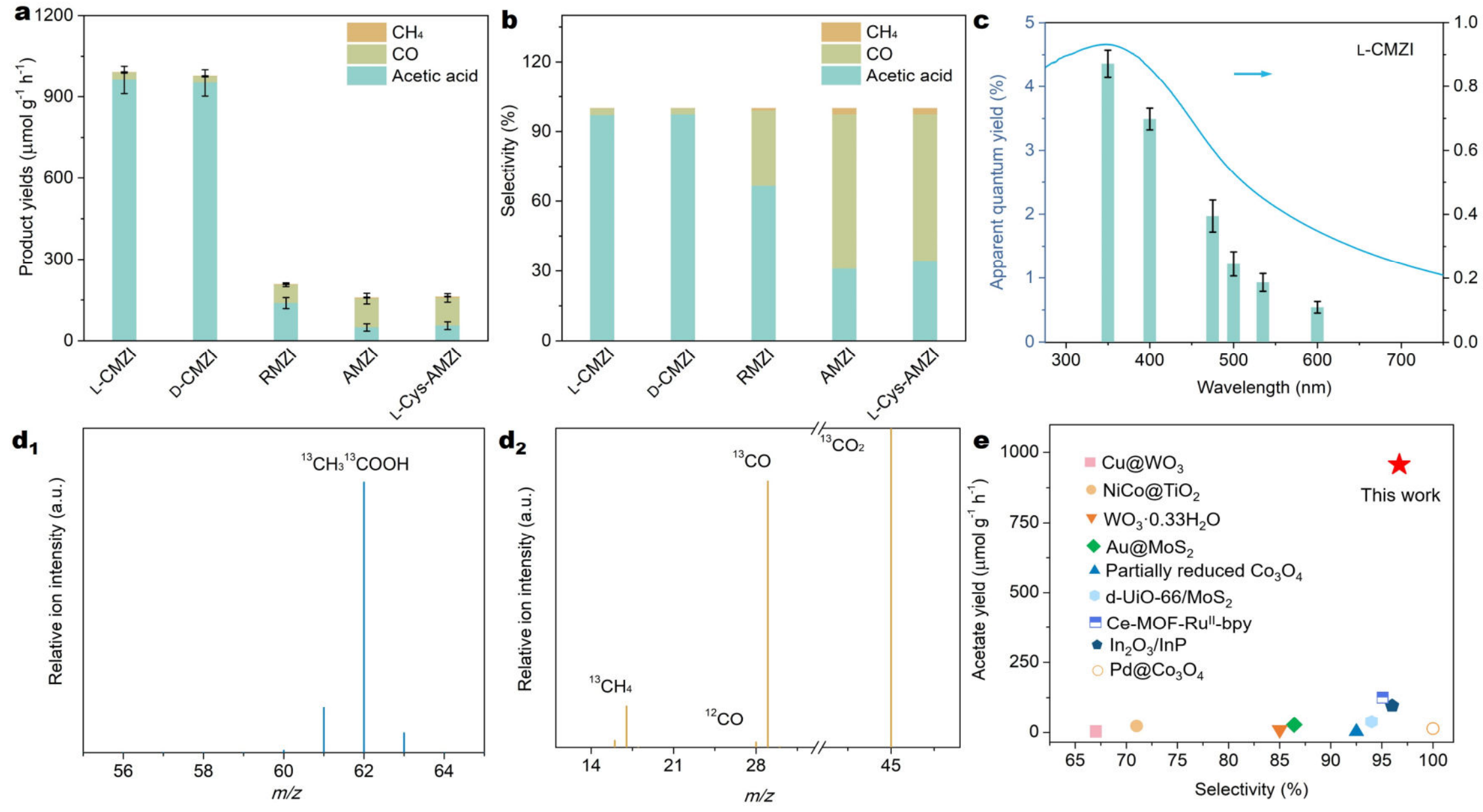

**Fig. 3 | PCCR performance on antipodal CMZIs, RMZI, AMZI and L-Cys-AMZI. a, b,** Production yields (**a**) and corresponding selectivity (**b**). **c,** AQY values of $CO_2$ photoreduction into acetic acid. **d$_{1,2}$,** GC-MS spectra of the products from photocatalytic $^{13}CO_2$ reduction on L-CMZI. **e,** Comparison for the evolution rate and selectivity of acetic acid between the CMZI and previously reported photocatalysts. The error bars represent the standard deviations of five independent measurements of the same sample.

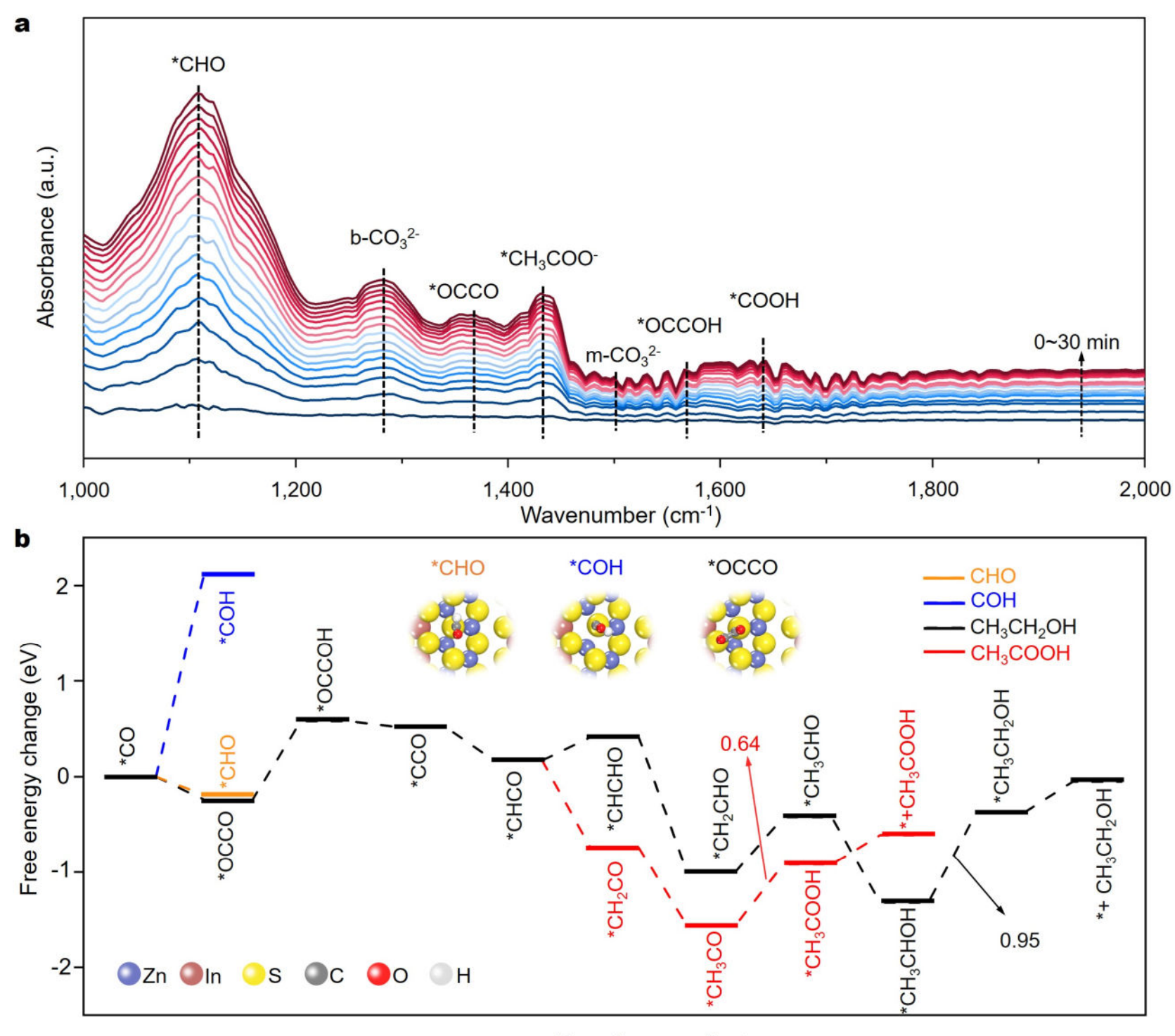

**Fig. 4 | Mechanistic studies of $CO_2$ reduction on CMZI. a,** In situ FTIR spectra of PCCR on L-CMZI, conducted over time under Xe lamp, equipped with an AM 1.5 G filter light irradiation. **b,** Calculated Gibbs free energy profiles for *CO reduction to acetate or ethanol on the $ZnIn_2S_4$(102) surface; insert shows the calculated adsorption structures (top view) of the key intermediate *CHO, *HCO, and *OCCO on the $ZnIn_2S_4$(102) surface.

## Theoretical calculations for the mechanism of preferential formation of acetic acid

Density functional theory (DFT) calculations were performed to elucidate the effect of CMZI on the formation of acetic acid (Fig. 4b). Considering $C_2H_5OH$ is the main possible byproducts from PCCR[37,45-47], the selectivity between $CH_3COOH$ and $C_2H_5OH$ formation was calculated. Models of $ZnIn_2S_4$(001) and (102) surfaces were constructed (Supplementary Figs. 29 and 30), which were observed to be mainly exposed on CMZIs. Given that *CO adsorption and activation constitute the key initial step in $CO_2$ reduction to ethanol and acetic acid, we evaluated these processes on $ZnIn_2S_4$(001)-$In_T$ and $ZnIn_2S_4$(102) surfaces. On $ZnIn_2S_4$(001)-$In_T$, CO adsorption is negligible across all investigated sites (Supplementary Fig. S31), with adsorption energies of −0.15, −0.14, and 0.01 eV. In contrast, CO adsorption at the In, Zn, and S sites of the $ZnIn_2S_4$(102) surface (Supplementary Fig. 32) is exothermic by 0.21, 0.24, and 0.97 eV, respectively. These results clearly indicate that $CO_2$ reduction predominantly proceeds on the $ZnIn_2S_4$ (102) surface, in agreement with prior reports[48,49].

Bader charge analysis further revealed weak CO adsorption at Zn and In sites, whereas a significantly stronger interaction occurs at the S site (Supplementary Table 6). These findings suggested that the S site acts as the predominant active center for subsequent hydrogenation.

Possible pathways for CO hydrogenation and C-C coupling were investigated (Fig. 4b and Supplementary Table 7). The calculations revealed that hydrogenating *CO to *COH is endothermic (2.13 eV), while *CHO formation is exothermic (−0.19 eV). Notably, the C-C coupling between two *CO species to generate the key *OCCO is even more favorable, with an exothermic energy of 0.24 eV, indicating that CO dimerization *via* C-C coupling indeed dominants and promotes $C_2$ products formation (e.g., $CH_3CH_2OH$ and $CH_3COOH$). To clarify the selectivity toward $CH_3COOH$ versus $CH_3CH_2OH$, we identified *CHCO as a key intermediate. Further hydrogenation to *CHCHO is endothermic (0.24 eV), whereas forming *$CH_2CO$ is strongly exothermic (−0.91 eV), suggesting a clear preference for the acetic acid pathway.

The results demonstrated that the intermediates along the $CH_3COOH$ pathway are consistently more stable than $CH_3CH_2OH$. The rate-determining step (RDS) for $CH_3CH_2OH$ formation is the hydrogenation of *$CH_3CHOH$ to *$CH_3CH_2OH$, which is endothermic by 0.95 eV. In contrast, the RDS for $CH_3COOH$ formation is the hydrogenation of *$CH_3CO$ to *$CH_3COOH$, which costs only 0.64 eV. These results showed that $CH_3COOH$ formation is both thermodynamically and kinetically favored over $CH_3CH_2OH$ on the $ZnIn_2S_4$(102) surface, consistent with the experimental results.

## Conclusions

In summary, CMZI establishes a new benchmark for PCCR by delivering unprecedented acetic acid yield and selectivity. It was revealed that spin polarization promotes the formation and stabilization of $^3$OCCO. The S sites play an essential role in steering the reaction pathway toward acetic acid by facilitating the RDS involving the hydrogenation of *$CH_3CO$ to *$CH_3COOH$. The spin polarization and S sites synergistically enable efficient C–C coupling and the consequent highly selective reaction pathway for acetic acid formation. Future investigations should leverage both the intrinsic properties of materials and the spin polarization effect to enable the synthesis of a broader range of scalable multicarbon products—such as ethylene glycol, oxalic acid, and propionic acid—*via* $CO_2$ reduction, as various catalysts can be engineered into chiral inorganic mesostructures. This strategy is also anticipated to diversify the range of reactions that involve triplet intermediates.

## METHODS：

### Synthesis of CMZI, RMZI and AMZI

All samples were synthesized by a facile hydrothermal method. In a typical synthesis of L-CMZI, $InCl_3 \cdot 4H_2O$ (0.147 g, 0.5 mmol), $ZnAc_2 \cdot 2H_2O$ (0.05 g, 0.25 mmol) and L-Cys (0.242 g, 2 mmol) were dissolved in deionized water. Then, 0.9 mL ethanolamine was added dropwise to the above solution while continuously stirring. The resulting solution 30 mL was then transferred into a 50 mL Teflon-lined stainless-steel autoclave. It was then heated to 150 °C and maintained at this temperature for a reaction period of 10 h. After cooling to room temperature, the samples were collected by centrifugation, washed with ultrapure water and ethanol several times, and dried in a vacuum oven at 50 °C. The syntheses of RMZI and AMZI are similar to that of CMZI, in which racemic Cys (0.242 g, 2 mmol) and TAA (0.15 g, 2 mmol) were used as a sulfide source instead of L-Cys, respectively.

### FTIR

FTIR spectra were collected using powder samples with a PerkinElmer FTIR Spectrometer.

### HPLC-MS

HPLC-MS analysis was performed by using an Acquity UPLC & XEVO G2-XS QTOF (Waters, United States) equipped with a UPLC column (Acquity UPLC HSS T3 1.8 μm, 2.1*100 mm). The detection of samples was conducted in positive ESI mode, and the function type was TOF-MS with a mass range of 50-600. Then, 1 μL of sample was injected and eluted using mobile phase A (0.1% mass concentration of formic acid aqueous solution) and mobile phase B (acetonitrile) at a flow rate of 0.35 mL/min.

### TGA

The thermogravimetric analysis (TGA) measurements were performed on a HITACHI STA200 simultaneous thermal analyzer with a heating rate of 10 ºC/min from room temperature to 750 ºC at Ar atmosphere using ~10 mg catalyst powder.

### XRD

The crystal structure of CMZI, RMZI and AMZI was examined by powder XRD, which was collected on a Rigaku MiniFlex 600 X-ray diffractometer equipped with Cu Kα (λ=1.5418 Å) radiation working at an acceleration voltage of 40 kV and a current of 80 mA. The scanning rate was set to 0.05° $s^{-1}$.

### SEM

Low-magnification SEM images were obtained using a JEOL JSM-7800 with an accelerating voltage of 5.0 kV.

### TEM/HRTEM and ET reconstruction

TEM images and SAED patterns of the CMZIs, RMZI and AMZI samples were acquired using JEOL JEM-F200 transmission electron microscope (high resolution pole-piece) equipped with a Schottky thermal field emission electron gun and working at 200 kV (Cs 1.0 mm, Cc 1.1 mm, point resolution of 1.9 Å for TEM mode) using a Gatan OneView IS camera (4096 ×4096 pixels). The RMZIs and AMZIs samples were gently crushed in a mortar and dispersed in ethanol with ultrasound, then dropped onto an ultrathin carbon film copper grid. The CMZIs samples were prepared by embedding and ultrathin sectioning into 100 nm thickness, then put onto an ultrathin carbon film copper grid. For zone-axis HRTEM imaging, a double tilt sample holder was used. The FFTs of the images were generated using a crystallographic image-processing software CRISP.

For 3D electron tomography, tilt series were captured using a high-tilt sample holder. A tilt series of 38 images for L-CMZI was captured from -44.24 º to +59.24 º at an interval of ~3 º, and a tilt series of 24 images for D-CMZI was captured from -50.07 º to +38.89 º at an interval of ~4 º. Images in the tilt series were preprocessed by removing the background, then aligned using center-of-mass (CoM) method and tilt-axis auto alignment. The alignment was further adjusted manually to reduce offset. Aligned image stacks were cropped to remove the blank area and then rescaled to the resolution of 375×375 pixels. Finally, the REal Space Iterative Reconstruction algorithm (RESIRE) was applied to the image stacks to generate the 375×375×375 voxels 3D tomograms. The cross section and volume rendering images of the tomograms were visualized by Tomviz software (https://tomviz.org).

### XPS

XPS was performed on an ESXCALAB Xi+ spectrometer (Thermo Fisher Corp.) with a 12 kV Al-Kα X-ray source with a step size of 0.1 eV. All the spectra were calibrated to the C 1s peak at 284.6 eV.

### Room temperature EPR

EPR measurements were conducted on powder samples at room temperature using a Bruker X-band (9.4 GHz) EMX plus 10/12 spectrometer equipped with an Oxford-910 cryostat and ITC-503 temperature controller (Oxford Instruments Ltd., Oxfordshire, UK).

### Specific surface area

The nitrogen adsorption/desorption isotherms were measured at 77 K with a Quantachrome QDI-SI. The surface area was calculated by the BET method using the desorption branch of the isotherm.

### UV-vis DRS

The UV-vis DRS were collected using powdered samples with a Lamda 950 UV-vis spectrophotometer, scanning at a rate of 50 nm $min^{-1}$ over a range of 250 to 800 nm.

The various bandgap energies ($E_g$) of samples were calculated according to the formula:

$$[\alpha h\nu]^n = A(h\nu - E_g) \quad (1)$$

where n = 2, representing $ZnIn_2S_4$, is a direct bandgap semiconductor. $\alpha$ and $\nu$ are the absorbance coefficient and light frequency, respectively.

### mc-AFM

mc-AFM current-voltage (I-V) measurements were performed using multimode AFM with Nanoscope V controller (Bruker-Dimension icon). I-V spectroscopy measurements were recorded by performing a voltage bias of −2.3 to +2.3 V at the tip in a contact mode. A magnetic Pt-coated Cr tip (Multi75E-G, Budget Sensors) with nominal spring constant 3 N $m^{-1}$ was used to acquire I-V curves. The tips are magnetized using a permanent magnet.

The sample is in a solid state, as used for photocatalysts, which were coated on a conductive substrate Ag film. In a typical measurement, a suspension was obtained by dispersing 5.0 mg of L-CMZI into a 1.0 mL solution containing water and ethanol (V/V=1/4). 0.1 mL of the suspension was sprayed onto the Ag film (1.0×10.0×10.0 mm). In this manner, a sample layer with a thickness of ~1-2 μm can be achieved.

The positions on the samples were picked randomly. During the conductive measurement, the magnetized tip vertically gets close to the sample surfaces, and the conductivity was measured with varying the external voltage. After completing a measurement, the tip was vertically elevated and move to another point to start a new measurement. To obtain a relatively accurate results, 30 points of each sample have been picked for the average performance.

### DRUV-vis and DRCD

DRUV-vis and DRCD spectra were obtained on a JASCO J-1500 spectropolarimeter fitted with a DRCD apparatus using powder samples, and data were collected with a scanning rate of 20 nm $min^{-1}$ ranging from 190 to 800 nm at 293 K.

The artifacts, such as the linear dichroism (LD) of all CD spectra, have been eliminated by the accumulation of data generated at 8 rotation degrees in the range of 360°.

### TAS and CPTAS

Femtosecond TAS experiments were carried out with a commercial spectrometer (Helios fire-EOS, Ultrafast Systems). The fundamental pulse was generated from a Ti: sapphire regenerative amplifier (Astrella, Coherent Inc., 800 nm, 100 fs, 7 mJ/pulse). A fraction of the fundamental beam was used to produce pump beams *via* an optical parametric amplifier (OPerA Solo, Coherent Inc.). A white light continuum (WLC) probe beam was generated by focusing the fundamental beam into a $CaF_2$ crystal, and the time window limit was 8 ns. The polarization between pump and probe pulses was set to be 54.7° during TA measurements. In CPTAS, L-CP and R-CP probe pulses were generated by using a quarter wave plate. 20 mg sample was dispersed in 1 mL of water and ultrasonically dispersed for 3 h.

**Transient photocurrent and EIS**

Transient photocurrent and EIS were measured in a conventional three-electrode process on an electrochemical workstation (CHI760E). Typically, a suspension was obtained by dispersing 5 mg samples into a 1 mL solution containing water and ethanol (V/V=1/4). The FTO conducting glass (1.1×10×20 mm) coating 0.1 mL of the suspension was the working electrode. Ag/AgCl and Platinum wire electrode were the reference electrode, and counter electrode, respectively. 0.5 M KCl aqueous solution served as the electrolyte.

**$CO_2$ photoreduction and products analysis**

The $CO_2$ photoreduction was carried out in a closed system using a quartz glass reactor with an approximate volume of 100 mL at 25 °C. Firstly, 5 mg of catalyst was dispersed into 50 mL of ultrapure water. 0.4 mg of $Na_2S$ and 2.9 mg of $Na_2SO_3$ were added to the dispersion and the mixture was sonicated for another 30 minutes to obtain homogeneity. Following this, the dispersion was first purged with $N_2$ to remove the dissolved air and then pure $CO_2$ gas was purged for 30 min to saturate the solution. Subsequently, the reactor was charged with ultra-pure $CO_2$ (greater than 99.999% purity) maintained at atmospheric pressure.

A 300 W Xe lamp (Beijing Ceaulight, CEL-HXF300-T3) equipped with an AM 1.5G filter sticks close to the window of the reactor as a light source, and the average light intensity irradiated on the photoreactor was ~100 mW $cm^{-2}$. The photocatalytic reactions were conducted over a period of 3 h. Lastly, the gas was collected in an aluminum bag and then injected into the gas chromatograph instrument GC9720Plus (equipped with a TCD and an FID detector, Fuli) for quantitative analysis of the products, where $CH_4$, CO and $CO_2$ were detected.

Liquid products were detected by $^1H$ NMR spectroscopy (Bruker Advance-III 500). 100 μL $D_2O$ and 450 μL of photocatalytic solution was mixed with 50 μL of 20 ppm dimethyl sulfoxide (DMSO) solution as internal standards for the $^1H$ NMR analysis. The concentration of acetic acid was calculated using the ratio of the area of the acetic acid peak (at a chemical shift of ~1.9 ppm) to that of the DMSO internal standard.

Catalytic performance was described by production yield calculated *via* equations (2 and 3).

Gas production yield ($Y_g$, μmol $g^{-1}$ $h^{-1}$),

$$Y_g = n_g/t \cdot m_{cat} \quad (2)$$

where $n_g$ represents the production of a molecule in gas, obtained from $n_g = pV_g/RT$. $p$ and $V_g$ are the total pressure and partial volume of a molecule in gas phase. $V_g$ was estimated by the GC report. The $m_{cat}$ stand for mass of photocatalyst.

Liquid production yield ($Y_l$, μmol $g^{-1}$ $h^{-1}$),

$$Y_l = n_l/t \cdot m_{cat} \quad (3)$$

where $n_l$ represents the production of a molecule in liquid phase, which was estimated from the report of $^1H$ NMR using DMSO (0.01408 μmol dissolved in 600 μL measured liquid) as an internal standard substance. The $t$ stand for reaction time.

Selectivity is defined as the ratio of the yield of the acetic acid to the sum of the yields of all products

$$\text{Selectivity} = \frac{\text{Yield of target product (μmol)}}{\text{Yields of all products (μmol)}} \quad (4)$$

**AQY**

The wavelength dependent AQY of CMZI for photocatalytic $CO_2$ reduction reaction was measured with different light band pass filter filters. After 3 h of $CO_2$ reduction the AQY was estimated from the following equation:

$$\text{AQY (\%)} = \frac{n \text{ (number of the elections taking part in reduction)}}{N \text{ (number of incident photons)}} \times 100\% \quad (5)$$

Number of reacted electrons were calculated from the yield of $CO_2$ reduced products. $n$ (acetic acid) = 8 × the yields of acetic acid × $N_A$, where is $N_A$ is Avogadro's number.

Number of incident photons, $N$, are calculated from the following equation:

$$N = \frac{PS\lambda t}{hc} \quad (6)$$

where, $P$ is the power density of the incident monochromatic light (W $m^{-2}$), $S$ ($m^2$) is the irradiation area, $t$ (s) is the duration of the incident light exposure and $\lambda$ (m) is the wavelength of the incident monochromatic light.

**Isotope labelling**

For isotope labelling $^{13}CO_2$ photoreduction experiment, the feeding gas was replaced with $^{13}CO_2$, other and conditions are the same as those for unlabeled $CO_2$ as mentioned above. The gas was collected in an aluminum bag and then injected into the GC-MS instrument 7890A-5975C (equipped with a thermal desorption detector, Agilent) for quantitative analysis of the products, where $^{13}CH_4$, $^{13}CO$ and $^{13}CO_2$ were detected.

The 10 μL liquid sample was analyzed using an GC-MS instrument 7890B-5977B (equipped with a headspace solid phase microextraction unit, Agilent) and placed in a headspace vial prior to injection.

**In situ DRIFTS**

DRIFTS measurements were conducted in situ. In brief, a Harrick Scientific HVCDRP reaction chamber coupled with a Praying Mantis DRIFTS accessory was employed. The base of the reaction cell was surrounded by a coil for cooling water circulation. The sample cup was set in the center of the cell, which had three windows on its dome. Two domes (KBr type) were used for the transmission of infrared radiation and the left dome (quartz type) was used for light introduction from a 300 W Xe lamp (Beijing Ceaulight, CEL-HXF300-T3) equipped with a 1.5 AM filter sticks close to the window of the reactor as a light source, and the average light intensity irradiated on the photoreactor was ~100 mW $cm^{-2}$.

**Low temperature EPR**

Low-temperature EPR measurements were conducted using a Bruker X-band (9.4 GHz) EMX plus 10/12 spectrometer equipped with an Oxford-910 cryostat and ITC-503 temperature controller (Oxford Instruments Ltd., Oxfordshire, UK). A cylindrical resonator (ER4119hs E011) was used for data collection. Samples were prepared and placed into quartz EPR tubes (Wilmad, 707-SQ-250 M, 4 mm OD) under atmosphere of $CO/CO_2$=5/95 (molar ratio). The samples were irradiated with a 300 W Xe lamp (Beijing Ceaulight, CEL-HXF300-T3) with a 1.5 AM filter and in situ EPR spectra were collected. The specific parameters for low-temperature EPR experiments were as follows: temperature 2 K, microwave power 0.1 mW, microwave frequency 9.410 GHz, modulation amplitude 0.5 mT, modulation frequency 100 kHz, and time constant 163.84 ms. For each sample, typically, 15 scans were taken to achieve a good signal-to-noise ratio.

**Computational details**

All structural optimization and total energy calculations in this work were performed at the spin-polarized DFT level by using the Vienna Ab initio Simulation Package (VASP)[50]. The projector augmented wave (PAW) method[51,52] and the Perdew-Burke-Ernzerh (PBE) functional under the generalized gradient approximation (GGA)[53,54] were applied throughout the calculations, and a cut-off energy of 400 eV was used. To avoid interactions between the neighboring slabs, a vacuum layer of ~15 Å was set. For the calculated models, the p (3×1) surface slab of $ZnIn_2S_4$(102) was constructed, and the surface slab contains four atomic layers with the bottom atomic layer being kept fixed. The Brillouin zones were sampled with k-point meshes of 2×2×1 Monk horst grids was applied to the system. The convergence criteria of energy and force in the calculations were set as $10^{-5}$ eV and 0.05 eV/Å, respectively.

The adsorption energies of different species (X) on the surface ($E_{ads}$(X)) were calculated with:

$$E_{ads}(X) = -(E_{X/slab} - E_{slab} - E_X) \quad (7)$$

where $E_{X/slab}$ represents the total energy of the adsorption system, while $E_{slab}$ and $E_X$ are the calculated energies of the slab and the gas-phase molecular X, respectively.

The Gibbs free energy G of each elementary step in the $CO_2$ reduction to acetate process was calculated as follows:

$$G = E + ZPE - T \times S \quad (8)$$

where E is the reaction energy calculated by DFT, and the changes of zero-point energy (ZPE) and entropy contribution (T × S) were obtained by calculating the frequency of the related system, and we set the reaction temperature as 298.15 K and 1 atm. The ZPE was calculated as follows[55]:

$$ZPE = \Sigma h\nu / 2 \quad (9)$$

The partition function ($pf$) and entropy ($S$) were calculated as follows[55]:

$$x_i = \frac{100ch\nu_i}{k_B T} \quad (10)$$

$$pf_i = \frac{x_i}{e^{x_i - 1}} - \log(1 - e^{-x_i}) \quad (11)$$

$$S = \frac{\Sigma(R \times pf_i)}{1000 \times 96.485} \quad (12)$$

where $c$, $h$, $v$, $k_B$, $T$ and $R$ are the light speed ($m \times s^{-1}$), Plank constant, frequency ($cm^{-1}$), Boltzman constant, temperature (298.15 K) and gas constant, respectively. The units of ZPE and $S$ are eV and eV/K, respectively. i is the number of vibration frequencies, and each adsorbed molecule has 3n vibration frequencies (n is the number of atoms). In addition, the frequencies of small gas phase molecules ($H_2$, CO and $H_2O$) were calculated using the Gaussian 16 program package[56] and 6-311G* basis sets[57] were used to obtain their ZPE and $S$ values.

## Acknowledgements

This work was supported by the Shanghai Natural Science Foundation (Grant No. 25ZR1402587, Y. F.), the Shanghai Titan Natural Science Development Foundation (Grant No. FF0010561, Y. F.), and the National Natural Science Foundation of China (Grant No. 22425303, L. H.; 92156024, 92356307, J. C.).

## Author Contributions

Y. F. conceived the idea and led the project. Y. C. synthesized the materials, performed the SEM, CD, XRD, electrochemical measurements, $CO_2$ photoreduction and products analysis; L. H., Y. L. and X. Z. worked on the structural characterization through HRTEM; J. C. and X. W. contributed to the measurement and analysis of lifetime and spin selectivity of materials; C. T., A. L. and L. Y. contributed to the EPR analysis; X. G. and Z. W. contributed to the theoretical calculations; Y. F., W. Z. and Y. C. contributed to the analysis of reaction mechanism and the preparation of manuscript.

## Competing interests

The authors declare no competing interests.

**Additional Information** is available in the online version of the paper.

## Data availability

The data that support the findings of this study are available from the corresponding authors upon request. Source data are provided in this paper.

**Correspondence and requests for materials** should be addressed to Yuxi Fang.